\documentclass[aps,pra,twocolumn,showpacs,groupedaddress]{revtex4}
\usepackage{graphicx,graphics,times}
\begin{document}
\title{Entanglement-assisted Enhanced Information Transmission Over a Quantum Channel with Correlated
Noise; A General Expression}
\author{A. Fahmi }
\email{ fahmi@theory.ipm.ac.ir}


\affiliation{ Institute for Studies in Theoretical Physics and
Mathematics (IPM) P. O. Box 19395-5531, Tehran, Iran}

\begin{abstract}
Entanglement and entanglement-assisted are useful resources to
enhance the mutual information of the Pauli channels, when the
noise on consecutive uses of the channel has some partial
correlations. In this paper, We study quantum communication
channels with correlated noise and derive a general expression for
the mutual information of quantum channel, for the product,
maximally entangled state coding and entanglement-assisted systems
with correlated noise in the Pauli quantum channels. Hence, we
suggest more efficient coding in the entanglement-assisted systems
for the transmission of classical information and derive a general
expression for the entanglement-assisted classical capacity. Our
results show that in the presence of memory, a higher amount of
classical information is transmitted by two or four consecutive
uses of entanglement-assisted systems.
\end{abstract}
 \maketitle
\section{INTRODUCTION}
One of the remarkable byproducts of the development of quantum
mechanics in recent years is quantum information and quantum
computation theories. Classical and quantum information theories
have some basic differences. Some of these differences are
superposition principle, uncertainty principle and non-local
effects. The non-locality associated with entanglement in quantum
mechanics is one of the most subtle and intriguing phenomena in
nature \cite{nil}. Its potential usefulness has been demonstrated
in a variety of applications, such as quantum teleportation,
quantum cryptography, and quantum dense coding. On the other hand,
quantum entanglement is a fragile feature, which can be destroyed
by interaction with the environment. This effect, which is due to
decoherence \cite{Shor},is the main obstacle for practical
implementation of quantum computing and quantum communication.
Several strategies have been devised against decoherence. Quantum
error correction codes, fault-tolerant quantum computation
\cite{nil} and dechorence free subspaces \cite{Knill} are among
them. One of the main problems in the quantum communication is the
decoherence effects in the quantum channels.

Recently, the study of quantum channels has attracted a lot of
attention. Early works in this direction were devoted, mainly, to
memoryless channels for which consecutive signal transmissions
through the channel are not correlated. The capacities of some of
these channels were determined \cite{Sch,Caves} and it was proven
that in most cases their capacities are additive for single uses
of the channel. For Gaussian channels under Gaussian inputs, the
multiplicativity of output purities was proven in \cite{Ser} and
the additivity of the energy-constrained capacity, even in the
presence of classical noise and thermal noise, was proven in
\cite{Hir}, under the assumption that successive uses of the
channel are represented by the tensor product of the operators
representing one single use of the channel, i.e., the channel is
memoryless.

In a recent letter, Bartlett \emph{et al.} \cite{Bar} showed that
it is possible to communicate with perfect fidelity, and without a
shared reference frame, at a rate that asymptotically approaches
one encoded qubit per transmitted qubit. They proposed a method to
encode a qubit, using photons in a decoherence-free subspace of
the collective noise model. Boileau \emph{et al.} considered
collective-noise channel effects in the quantum key distribution
\cite{Boi1} and they gave a realistic roboust scheme for quantum
communication, with polarized entangled photon pairs \cite{Boi2}.
In the last few years much attention has been given to bosonic
quantum channels \cite{Gio}.

Recently Macchiavello \emph{et al.} \cite{Mac,Bowen}, considered a
different class of channels, in which correlated noise acts on
consecutive uses of channels. They showed that higher mutual
information can be achieved above a certain memory threshold, by
entangling two consecutive uses of the channel. This type of
channels and its extension to the bosonic case, has attracted a
lot of attention in the recent years \cite{Gio2}.

K. Banaszek \emph{et. al} \cite{Ball} implemented Macchiavello
\emph{et al.} suggestion experimentally. They shown how
entanglement can be used to enhance classical communication over a
noisy channel. In their setting, the introduction of entanglement
between two photons is required in order to maximize the amount of
information that can be encoded in their joint polarization degree
of freedom, and they obtained experimental classical capacity with
entangled states and showed that it is more than $2.5$ times the
theoretical upper limit, when no quantum correlations are allowed.
Hence, recently some people show that provided the sender and
receiver share prior entanglement, a higher amount of classical
information is transmitted over Pauli channels in the presence of
memory, as compared to product and entangled state coding
\cite{AT}.

In this Paper, we show that if parties use a semi-quantum approach
to entanglement-assisted coding, a higher amount of classical
information is transmitted. We derive a general expression for the
entanglement-assisted classical capacity and compare our results
with the product Bell states and entanglement-assisted coding for
various types of Pauli channels.

This Paper is organized as follows: In Sec. II we briefly review
some properties of quantum memory channels and derive the general
expression of classical capacity of quantum channel for product
and maximally entangled state coding. In Sec. III, we derive the
general expression of entanglement-assisted classical capacity of
quantum channels with correlated noise that was calculated
previously for Pauli channels by Arshed and Toor (AT). In Sec IV,
we show that AT model is not an optimal coding for transmission of
classical information, and we suggest another sets of states and
derive the general expression of entanglement-assisted classical
capacity. Our results show that a higher amount of classical
capacity can indeed be achieved for all values of memory, by two
or four consecutive uses of the entanglement-assisted systems for
depolarizing, flip, two Pauli and phase damping channels.

\section{Entanglement-enhanced information transmission over a quantum channel with correlated noise }

Encoding classical information into quantum states of physical
systems gives a physical implementation of the constructs of
information theory. The majority of research into quantum
communication channels has focused on the memoryless case,
although there have been a number of important results obtained
for quantum channels with correlated noise operators or more
general quantum channels \cite{Mac,Ha}.

The action of transmission channels is described by Kraus
operators $A_{i}$ \cite{Kraus}, which satisfy the
$\sum_{i}A^{\dagger}_{i}A_{i}\leq1$, the equality holds when the
map is trace-preserving. Thus, if we send a qubit in a state
described by the density operator $\rho$, through the channel,
then the corresponding out put state is given by the map.
\begin{eqnarray*}
\rho \longrightarrow \varepsilon(\rho)=\sum_{i}A_{i} \rho
A^{\dagger}_{i}
\end{eqnarray*}

An interesting class of Kraus operators acting on individual
qubits  can be expressed in terms of the Pauli operators
$\sigma_{x,y,z}$
\begin{eqnarray*}
A_{i} = \sqrt{p_{i}} \sigma_{i}\ ,
\end{eqnarray*}
with $\sum_i p_i = 1$ , $i=0,x,y,z$ and $ \sigma_{0}=I$. A noise
model for these actions is, for instance, the application of a
random rotation by angle $\pi$ around the axis $
\hat{\bf{x}},\hat{\bf{y}},\hat{\bf{z}}$ with the probabilities
$p_x , p_y , p_z$ respectively, and the identity with probability
$p_0$.

In the simplest scenario, the transmitter can send one qubit at a
time along the channel. In  this case the codewords will be
restricted to the tensor products of the states of the individual
qubits. Quantum mechanics, however allows also the possibility to
entangle multiple uses of the channel.

Recently, a model for quantum channels with memory has been
proposed that can consistently define quantum channels with
Markovian correlated noise \cite{Mac}. The model is also extended
to describe channels that act on transmitted states in such a way
that there is no requirement for interactions with an environment
within the model. A Markovian correlated noise channel of length
$n$, is of the form:

\begin{eqnarray}
\varepsilon(\rho )&=&\sum_{i_1\cdots i_n}  p_{i_1}
p_{i_2|i_1}\dots p_{i_n|i_{n-1}}
\\\nonumber
&&\times(A_{i_n}\otimes \cdots \otimes A_{i_1}) \rho
(A^{\dagger}_{i_1}\otimes \cdots \otimes A^{\dagger}_{i_n})
\end{eqnarray}
where the $A_{i_k}$ are Kraus operators for single uses of the
channel on the state $k$ and $p_{i_k|i_{k-1}}$ can be interpreted
as the conditional probability that a $\pi$ rotation around the
axis $i_k$ is applied to the $k$-th qubit, given that a $\pi$
rotation around axis $i_{k-1}$ was applied on the  $k-1$-th qubit.
They considered \cite{Mac} the case of two consecutive uses of a
channel with a partial memory, i.e. $ p_{i_k|i_{k-1}} = (1-\mu )
p_{i_k} + \mu \delta_{i_k|i_{k-1}}$. This means that with the
probability $ \mu $, the same rotation is applied to both qubits,
while with probability $1 - \mu $ the two rotations are
uncorrelated. Then, they concentrated their attention on the
depolarizing channel, for which $p_0=1-p$ and $p_i=p/3, i=x,y,z$,
and showed that  for the specific case of a quantum depolarizing
channel with collective noise, the transmission of classical
information can be enhanced by employing maximally entangled
states as carriers of information, rather than product states.

In this section, we would like derive a general expression for the
classical capacity of Pauli quantum channels with correlated noise
for Bell and product states. The maximum mutual information
$I(\varepsilon (\rho))$ of a general quantum channel $\varepsilon$
is given by the Holevo-Schumacher-Westmoreland bound \cite{Sch}
\begin{eqnarray}\label{Sch}
C(\varepsilon)=Max_{[\pi_{i},\rho_{i}]} S(\varepsilon (\sum_{i}
\pi_{i}\rho_{i})) -\sum_{i}\pi_{i}S(\varepsilon(\rho_{i}))
\end{eqnarray}
where $S(\omega)=-Tr(\omega\log_{2}\omega)$ is the von Neumann
entropy of the density operator $\omega$ and the maximization is
performed over all input ensembles $\pi_{i}$ and $\rho_{i}$. Note
that this bound incorporates maximization over all POVM (positive
operator value measures) measurements at the receiver, including
the collective ones over multiple uses of the channel.

In what follows, we shall derive $I(\varepsilon (\rho))$ for
maximally Bell and product states. For the Bell states which are
defined as:

\begin{eqnarray}\label{Bell}
|\Psi^{\pm}\rangle_{AB} &=&\frac{1}{\sqrt 2}(|00\rangle \pm
|11\rangle),
\nonumber\\
|\Phi^{\pm}\rangle_{AB} &=&\frac{1}{\sqrt 2} (|01\rangle \pm
|10\rangle)
\end{eqnarray}
we know that the maximally entangled input states can be derived
from each other by unitary transformations (for example
$\rho_{0}=|\Psi^{+}\rangle\langle\Psi^{+}|$), and with respect to
the von Neumann entropy, $S(\omega)$ is invariant under any
unitary transformation of a quantum state $\omega$. The second
term on the right hand side of equation (\ref{Sch}) becomes:
\begin{eqnarray*}
\sum_{i}\pi_{i}S(\varepsilon(\rho_{i}))=
S(\varepsilon(\rho_{0}))=-\sum_{i}\lambda_{i}\log_{2}\lambda_{i}
\end{eqnarray*}
where $\lambda_{i}$ are eigenvalues of the transformed states. On
the other hand, with respect to the relation
$Tr_{k}\rho_{i}=\frac{1}{2}\textbf{I}_{l}$ (with $k=A,B$ and
$l=B,A$), we show that the Holevo limit can be attained by setting
$\pi_{i}=\frac{1}{2^{2}}$ (with $i=0,...,3$). The quantum state
$\varepsilon(\rho_{i})$, in the first term, can be writhen as:

\begin{eqnarray*}
\varepsilon (\sum_{i} \pi_{i}\rho_{i})=
\frac{1}{2^{2}}\sum_{i}\varepsilon(\rho_{i})
=\varepsilon(\frac{1}{2}\textbf{I}_{1}\otimes\frac{1}{2}\textbf{I}_{2})
\end{eqnarray*}
To get the final result, we take suitable bases for the density
matrixes representation. Then, the mutual information
$I(\varepsilon (\rho))$ of quantum channel is given by:

\begin{eqnarray}\label{Mu}
I(\varepsilon(\rho))=2+\sum_{i}\lambda_{i}\log_{2}\lambda_{i}
\end{eqnarray}
which for Bell states have the eigenvalues:
\begin{eqnarray}
\lambda_{1}&=&(1-\mu)[(p_{0}^{2}+p_{x}^{2}+p_{y}^{2}+p_{z}^{2})+\mu\\\nonumber
\lambda_{2}&=&(1-\mu)2(p_{0}p_{z}+p_{x}p_{y})\\\nonumber
\lambda_{3}&=&(1-\mu)2(p_{0}p_{x}+p_{y}p_{z})\\\nonumber
\lambda_{4}&=&(1-\mu)2(p_{0}p_{y}+p_{x}p_{z})
\end{eqnarray}

With a similar approach, it is straightforward to verify that
eigenvalues of product states in the $z$ basis ${|jk\rangle_{z}}$
(with $j,k=0,1$) are give by:

\begin{eqnarray}
\lambda^{z}_{1}&=&(1-\mu)(p_{0}+p_{z})^{2}+\mu(p_{0}+p_{z})\\\nonumber
\lambda^{z}_{2}&=&(1-\mu)(p_{x}+p_{y})^{2}+\mu(p_{x}+p_{y})\\\nonumber
\lambda^{z}_{3,4}&=&(1-\mu)(p_{0}+p_{z})(p_{x}+p_{y})
\end{eqnarray}
By inserting these eigenvalues in the eq. (\ref{Mu}), the mutual
information is give for the Bell and product states.

It is not complicated to verify that if we use product states in
$x$ basis ${|jk\rangle_{x}}$ (with $j,k=0,1$ and
$|l\rangle_{x}=(|0\rangle+(-1)^{l}|\rangle)_{z}$, $l=j,k$), the
eigenvalues in this basis are identical to the above eigenvalues
with $p_{x}$ and $p_{z}$ being interchanged. Hence, if we use the
product states in the $y$ basis ${|jk\rangle_{y}}$ (with $j,k=0,1$
and $|l\rangle_{y}=(|0\rangle+i(-1)^{l}|\rangle)_{z}$, $l=j,k$),
the eigenvalues in this basis are identical to above eigenvalues
with, $p_{y}$ and $p_{z}$ being interchanged.

These eigenvalues reduce to previous amounts in the especial case
of depolarizing channels \cite{Mac}. In the final section, we
compare these mutual information to the entanglement-assisted
classical capacity.

\section{Entanglement-assisted Enhanced classical capacity of quantum channels with correlated noise}
Recently, Arshed and Toor \cite{AT} considered an interesting
extension of classical capacity, and calculated the
entanglement-assisted classical capacity of Pauli channels for two
consecutive uses of the channels. They assumed that the sender
(Alice) and the receiver (Bob) share two (same or different)
maximally entangled Bell states $|\psi^{\pm}\rangle,$
$|\phi^{\pm}\rangle$ (as defined by the eq. (\ref{Bell})). The
first qubits of Bell states belongs to Alice while the second
qubits belongs to Bob. Alice sends her qubits through the channel.
They calculated eigenvalues of the pure density matrix,
transformed under the action of the depolarizing channel, for
probabilities $p_{0}=1-p$, $p_{x}=p_{y}=p_{z}=p/3$, and after some
simple calculations, they derived the entanglement-assisted
classical capacity for the Pauli channels in the presence of
partial memory.

As was suggested by AT, in the presence of partial memory, the
action of Pauli channels is described by the Kraus operators:
\begin{eqnarray*}
A_{i,j}(\mu)=\sqrt{p_{i}[(1-\mu)p_{j}+\mu\delta_{i,j}]}
(\sigma_{i}^{A}\otimes I^{B})\otimes (\sigma_{j}^{A}\otimes I^{B})
\end{eqnarray*}
The mutual information of quantum dense coding system, where Alice
and Bob shared quantum state $\rho^{AB}$ (which is statistical
mixture of the Bell states), is given by:
\begin{eqnarray}
I^{AT}(\varepsilon(\rho))=4+\sum_{i}\lambda_{i}\log_{2}\lambda_{i}
\end{eqnarray}
where $\lambda_{i}$ are the eigenvalues of the transformed states.
For states that were considered by AT, eigenvalues in the general
case are give by:
\begin{eqnarray}
\lambda_{1}&=&(1-\mu)p_{0}^{2}+\mu p_{0}\\\nonumber
\lambda_{2}&=&(1-\mu)p_{x}^{2}+\mu p_{x}\\\nonumber
\lambda_{3}&=&(1-\mu)p_{y}^{2}+\mu p_{y}\\\nonumber
\lambda_{4}&=&(1-\mu)p_{z}^{2}+\mu p_{z}\\\nonumber
\lambda_{5,6}&=&(1-\mu)p_{0}p_{x}\\\nonumber
\lambda_{7,8}&=&(1-\mu)p_{0}p_{y}\\\nonumber
\lambda_{9,10}&=&(1-\mu)p_{0}p_{z}\\\nonumber
\lambda_{11,12}&=&(1-\mu)p_{x}p_{y}\\\nonumber
\lambda_{13,14}&=&(1-\mu)p_{x}p_{z}\\\nonumber
\lambda_{15,16}&=&(1-\mu)p_{y}p_{z}
\end{eqnarray}

These eigenvalues reduce those of AT approach in the especial
cases of depolarizing, flip, two-Pauli and phase damping channels
\cite{AT}.

\section{Semi-Entanglement-assisted Enhanced Information transmission over quantum channels with correlated noise}
In the preceding section, we derived the classical information
capacity $C(\varepsilon)$ of a quantum dense coding system.
Although, the use of entanglement-assisted enhances classical
capacity of channels with correlated noise, in comparison with
both product and entangled state coding for all values of $\mu$,
here exists other quantum dense coding systems in which those
states have a higher amount of classical information transmission
in the channels with correlated noise.

In this section, we derive a general expression for the
entanglement-assisted classical capacity for two kinds of quantum
dense coding systems which we call semi-quantum dense coding. Our
results show that for channels with partial memory and with the
use of appropriate choice of entanglement-assisted states, higher
mutual information can indeed be achieved for all values of $\mu$,
by two or four consecutive uses of the channel. First, we consider
a simple modification of AT approach. As we saw, AT use two
maximally entangled Bell states, the first qubit belongs to the
Alice and the second qubit belongs to the Bob, we replace AT
states by following states:
\begin{eqnarray}\label{Sem-Qun1}
|\Phi_{1,2}\rangle=|\psi^{+}\rangle_{A}|\psi^{+}\rangle_{B}\pm|\psi^{-}\rangle_{A}|\psi^{-}\rangle_{B}\\\nonumber
|\Phi_{3,4}\rangle=|\psi^{-}\rangle_{A}|\psi^{+}\rangle_{B}\pm|\psi^{+}\rangle_{A}|\psi^{-}\rangle_{B}\\\nonumber
|\Phi_{5,6}\rangle=|\varphi^{+}\rangle_{A}|\psi^{+}\rangle_{B}\pm|\varphi^{-}\rangle_{A}|\psi^{-}\rangle_{B}\\\nonumber
|\Phi_{7,8}\rangle=|\varphi^{-}\rangle_{A}|\psi^{+}\rangle_{B}\pm|\varphi^{+}\rangle_{A}|\psi^{-}\rangle_{B}\\\nonumber
|\Phi_{9,10}\rangle=|\psi^{+}\rangle_{A}|\varphi^{+}\rangle_{B}\pm|\psi^{-}\rangle_{A}|\varphi^{-}\rangle_{B}\\\nonumber
|\Phi_{11,12}\rangle=|\psi^{-}\rangle_{A}|\varphi^{+}\rangle_{B}\pm|\psi^{+}\rangle_{A}|\varphi^{-}\rangle_{B}\\\nonumber
|\Phi_{13,14}\rangle=|\varphi^{+}\rangle_{A}|\varphi^{+}\rangle_{B}\pm|\varphi^{-}\rangle_{A}|\varphi^{-}\rangle_{B}\\\nonumber
|\Phi_{15,16}\rangle=|\varphi^{-}\rangle_{A}|\varphi^{+}\rangle_{B}\pm|\varphi^{+}\rangle_{A}|\varphi^{-}\rangle_{B}
\end{eqnarray}
In these states, the first Bell states belongs to Alice and the
second Bell states belongs to Bob, which are represented by
subscriptions in the Bell states respectively. These states are
equivalent to the Bell state which is shared by Alice and Bob. For
example, parties can construct
$|\Phi_{1}\rangle=|00\rangle_{A}|00\rangle_{B}+|11\rangle_{A}|11\rangle_{B}$
by performing \textsc{cnot} gates $C_{A1}$ and $C_{B2}$ on the
state
$(|00\rangle_{AB}+|11\rangle_{AB})|0\rangle_{1}|0\rangle_{2}$ ($A$
and $B$ are the controller and $1$ and $2$ are the targets).
Hence, Alice can transform $|\Phi_{i}\rangle$ (with $i=1,...,8$)
states to each other by a local operation on the his qubits. The
same transformation exists for $|\Phi_{i}\rangle$ (with
$i=9,...,16$). These two groups of states cannot transform to each
other by Alice's local operation.

On the other hand, as we know, if Alice and Bob previously shared
entangled states between themselves, then, for noiseless quantum
channels, the amount of classical information transmitted through
a quantum channel is doubled in comparison with the unshared
states $C_{E}=2C$ \cite{BW}. In other words, every previously
shared Bell state is equivalent with two quantum channels (or
twice the using quantum channel).

In our approach, Alice and Bob shared only one of the above states
(for example, $|\Phi_{1}\rangle$) and they had previously
compromised that if Alice states were received by Bob in the time
interval $\delta t_{0}=t_{1}-t_{0}$ ($\delta t_{1}=t_{2}-t_{1}$),
Bob would operate $I^{B}\otimes I^{B}$ ($I^{B}\otimes
\sigma_{x}^{B}$) on the his qubits (the Bell states that were
represented by subscript $2$ in the (\ref{Sem-Qun1})) \cite{Ano1}.

By using this protocol, Alice and Bob get access to all Hilbert
space states in the states (\ref{Sem-Qun1}). This protocol (using
one shared Bell state between parties and two types of qubits
transmission), we call semi-quantum dense coding which has the
same cost as that of the AT approach (which uses two shared Bell
states between parties and one type of qubits transmission in the
quantum channel).

We consider the shortest variational time of the channel (fiber)
under thermal and mechanical fluctuations as $\tau_{fluc}$. If the
time lapse ($\tau_{lap}$ ) between the two used channel is small
compared to $\tau_{fluc}$, the effects of the channel on various
qubits can be considered as a correlated noise. For example, in
the experiment of K. Banaszek \emph{et. al} \cite{Ball}
$\tau_{lap}\approx 6 ns$, which is much smaller than the
mechanical fluctuations of the fiber.

Similar to the previous cases, the action of transmission channels
is described by Karus operators $A_{i,j}$ , satisfying
$\sum_{i,j}A_{i,j}A_{i,j}^{\dagger}=1$, Where in the presence of
partial memory, the action of Pauli channels on $\rho$ is
described by
\begin{eqnarray*}
A_{i,j}(\mu)=\sqrt{p_{i}[(1-\mu)p_{j}+\mu\delta_{i,j}]}
(\sigma_{i}\otimes\sigma_{j})\otimes (I\otimes I)
\end{eqnarray*}

If Alice sends the density matrix $\rho$ through the quantum
channel, the corresponding output state is given by the mapping
\begin{eqnarray}
\rho \longrightarrow
\varepsilon(\rho)=\sum_{i,j=0}^{3}A_{i,j}(\mu)\rho
A_{i,j}^{\dagger}(\mu)
\end{eqnarray}
That can be written for channels with partial memory as:
\begin{widetext}
\begin{eqnarray*}
\varepsilon(\rho)&=&(1-\mu)\sum_{i,j=0}^{3}p_{i}p_{j}
(\sigma_{i}\otimes\sigma_{j}\otimes I\otimes I)\rho
(\sigma_{i}\otimes\sigma_{j}\otimes I\otimes I)
+\mu\sum_{i=0}^{3}p_{i}(\sigma_{i}\otimes\sigma_{i}\otimes
I\otimes I)\rho (\sigma_{i}\otimes\sigma_{i}\otimes I\otimes I)
\end{eqnarray*}
\end{widetext}

Here, we would like to compute the entanglement-assisted classical
capacity of quantum channels by using quantum superdense coding
system, where Alice and Bob share a completely entangled state
$|\Phi^{AB}_{1}\rangle$, which belongs to an $(2^{2}\times
2^{2})$-dimensional (for $2^{2}$-dimensional Bell states) Hilbert
space $\emph{H}\otimes\emph{H}$, and the encoded system is sent
through a quantum channel, described by an arbitrary
trace-preserving completely positive map $\varepsilon$.

In a simple modification of Ban \emph{et al.} \cite{Ban} approach
to the calculation of classical capacity of quantum dense coding,
we assume Alice and Bob to share a statistical mixture of the
entangled states $|\Phi_{i}\rangle$ (with $i=1,...,16$), where the
shared quantum state $\rho^{AB}$ is given by:

\begin{eqnarray}
\rho^{AB}=\sum_{j=0}^{16}\lambda_{j}|\Phi^{AB}_{j}\rangle\langle\Phi^{AB}_{j}|
\end{eqnarray}
with $\lambda_{j}\geq0$ and $\sum_{j=0}^{16}\lambda_{j}=1$. To
encode $2\log_{2}2^{2}$ bits of classical information, Alice
applies one of $8\times2$ operation, $8$ unitary operators $V_{k}$
$(k=0,...,8)$ to her part of the quantum state $\rho^{AB}$ and the
two types of state transmission. She sends the encoded system to
Bob through an arbitrary quantum channel $\varepsilon$. Then, Bob
obtains the quantum state:

\begin{eqnarray}
\rho'^{AB}_{jk}=(\varepsilon^{A}\otimes I^{B})[(V^{A}_{j}\otimes
v^{B}_{k})\rho^{AB}(V^{A\dagger}_{j}\otimes v^{B\dagger}_{k})]
\end{eqnarray}
In the above relation, $v^{B}_{k}$ depends on the type of qubits
sent by Alice, which operate by Bob on the his qubits. If the
prior probability of the classical information corresponding to
$V^{A}_{j}$ and $v^{B}_{k}$ (Alice and Bob operation on the shared
state) is $\pi_{jk}$, where $\pi_{jk}\geq0$ and
$\sum_{j=0}^{8}\sum_{k=0}^{1}\pi_{jk}=1$, the maximum amount of
mutual information of the quantum dense coding system would be
given by \cite{BSST}:

\begin{eqnarray}\label{BS}
C^{E}(\varepsilon, {\pi_{jk}})&=&Max_{[\rho,
{\pi_{jk}}]}I^{AB}(\varepsilon,
{\pi_{jk}})\\\nonumber&=&Max_{[\rho, {\pi_{jk}}]} S(\rho'^{AB})
-\sum_{j,k}\pi_{jk}S(\varepsilon(\rho_{jk}))
\end{eqnarray}
where $\rho'^{AB}=\sum_{j}\sum_{k}\pi_{jk}\varepsilon
(\rho^{AB}_{jk})$. In the above relation the maximization is
performed over all input ensembles $\pi_{jk}$ and $\rho_{ik}$.
Note that this bound incorporates maximization over all POVM
 measurements (positive operator value measures) at the receiver,
including the collective ones over multiple uses of the channel.
In what follows, we shall derive $I(\varepsilon (\rho_{jk}))$ for
entangled states which were suggested in the eq.(\ref{Sem-Qun1}).
We know that the entangled input states $|\Phi_{i}\rangle$ can be
derived from $|\Phi_{1}\rangle$ by unitary transformations (Pauli
matrixes), and with respect to the von Neumann entropy,
$S(\omega)$ is invariant under any unitary transformation of a
quantum state $\omega$. The second term on the right hand side of
equation (\ref{BS}) becomes:
\begin{eqnarray*}
\sum_{j,k}\pi_{jk}S(\varepsilon(\rho_{jk}))=
S(\varepsilon^{A}\otimes\emph{I}^{B}(\rho^{AB}))
\end{eqnarray*}
On the other hand, we show that the maximum amount of mutual
information can be attained by setting $\pi_{jk}=\frac{1}{2^{4}}$
(with $j=0,...,8$ and $k=0,1$). The quantum state $\varepsilon
(\rho^{AB}_{jk})$, in the first term, can be writhen as:
\begin{eqnarray*}
\sum_{jk}\pi_{jk}\varepsilon(\rho^{AB}_{jk})=
\frac{1}{2^{4}}\sum_{jk}\varepsilon(\rho^{AB}_{jk})
=\varepsilon(\frac{1}{2}\textbf{I}_{1}\otimes...\otimes\frac{1}{2}\textbf{I}_{4})
\end{eqnarray*}
To get the final result, we take suitable bases for density
matrixes representation. Mutual information
$I_{1}^{AB,e-a}(\varepsilon, {\pi_{jk}})$ can be calculated for
the quantum channel with entangled states (\ref{Sem-Qun1}) in the
general case. Therefore, for symmetric and asymmetric Pauli
channels, we have:
\begin{eqnarray}
I_{1}^{AB,e-a}(\varepsilon, {\pi_{jk}})=
4-S(\varepsilon^{A}\otimes\emph{I}^{B}(\rho^{AB}))
\end{eqnarray}
The von Neumann entropy of the $\rho^{AB}$ state, transformed
under the action of Pauli channels, is given by:

\begin{eqnarray}
S(\varepsilon^{A}\otimes\emph{I}^{B}(\rho^{AB}))=-\sum_{i}\lambda_{i}\log_{2}\lambda_{i}
\end{eqnarray}
where $\lambda_{i}$ are the eigenvalues of the transformed
$\rho^{AB}$ state, which have the explicit forms:

\begin{eqnarray}
\lambda_{1}&=&(1-\mu)(p_{0}^{2}+p_{z}^{2})+\mu(p_{0}+p_{z})\\\nonumber
\lambda_{2}&=&(1-\mu)(p_{x}^{2}+p_{y}^{2})+\mu(p_{x}+p_{y})\\\nonumber
\lambda_{3}&=&(1-\mu)2p_{0}p_{z}\\\nonumber
\lambda_{4}&=&(1-\mu)2p_{x}p_{y}\\\nonumber
\lambda_{5,6}&=&(1-\mu)(p_{0}p_{x}+p_{y}p_{z})\\\nonumber
\lambda_{7,8}&=&(1-\mu)(p_{0}p_{y}+p_{x}p_{z})
\end{eqnarray}
Hence, there exist two groups of eigenvalues which are equal to
each other. These states don't mix with each other through
interaction with environment.

Although the above states are more efficient than the AT states
(as we shall see at the end of this section) for the transmission
of classical information, there exist other entangled states which
enhance the classical capacity of quantum channels through the use
of the entanglement-assisted. Similar to the previous case, we
consider semi-quantum approach, which uses one entangled state
augmented by quantum channels and show that in the channels with
partial memory, the use of four particle entanglement-assisted
enhances the amount of mutual information can indeed be achieved
for all values of $\mu$, by four consecutive uses of the channel
(at more of cases, for example, in the depolarizing, flip and two
Pauli channels). This approach has $8^{4}$ dimensional Hilbert
space, that in following we represent sixty four (one quarter of
total Hilbert space dimension) of them:

\begin{widetext}
\begin{eqnarray}\label{states}
|\Psi_{1,i}\rangle=|\psi^{+}\rangle_{1}|\psi^{+}\rangle_{2}|\omega^{i}\rangle_{3}|\psi^{+}\rangle_{4}
+|\psi^{-}\rangle_{1}|\psi^{-}\rangle_{2}|\omega^{i+1}\rangle_{3}|\psi^{-}\rangle_{4}
+|\phi^{+}\rangle_{1}|\phi^{+}\rangle_{2}|\omega^{i+2}\rangle_{3}|\phi^{+}\rangle_{4}
+|\phi^{-}\rangle_{1}|\phi^{-}\rangle_{2}|\omega^{i+3}\rangle_{3}|\phi^{-}\rangle_{4}\\\nonumber
|\Psi_{2,i}\rangle=|\psi^{+}\rangle_{1}|\psi^{+}\rangle_{2}|\omega^{i}\rangle_{3}|\psi^{+}\rangle_{4}
+|\psi^{-}\rangle_{1}|\psi^{-}\rangle_{2}|\omega^{i+1}\rangle_{3}|\psi^{-}\rangle_{4}
-|\phi^{+}\rangle_{1}|\phi^{+}\rangle_{2}|\omega^{i+2}\rangle_{3}|\phi^{+}\rangle_{4}
-|\phi^{-}\rangle_{1}|\phi^{-}\rangle_{2}|\omega^{i+3}\rangle_{3}|\phi^{-}\rangle_{4}\\\nonumber
|\Psi_{3,i}\rangle=|\psi^{+}\rangle_{1}|\psi^{+}\rangle_{2}|\omega^{i}\rangle_{3}|\psi^{+}\rangle_{4}
-|\psi^{-}\rangle_{1}|\psi^{-}\rangle_{2}|\omega^{i+1}\rangle_{3}|\psi^{-}\rangle_{4}
-|\phi^{+}\rangle_{1}|\phi^{+}\rangle_{2}|\omega^{i+2}\rangle_{3}|\phi^{+}\rangle_{4}
+|\phi^{-}\rangle_{1}|\phi^{-}\rangle_{2}|\omega^{i+3}\rangle_{3}|\phi^{-}\rangle_{4}\\\nonumber
|\Psi_{4,i}\rangle=|\psi^{+}\rangle_{1}|\psi^{+}\rangle_{2}|\omega^{i}\rangle_{3}|\psi^{+}\rangle_{4}
-|\psi^{-}\rangle_{1}|\psi^{-}\rangle_{2}|\omega^{i+1}\rangle_{3}|\psi^{-}\rangle_{4}
+|\phi^{+}\rangle_{1}|\phi^{+}\rangle_{2}|\omega^{i+2}\rangle_{3}|\phi^{+}\rangle_{4}
-|\phi^{-}\rangle_{1}|\phi^{-}\rangle_{2}|\omega^{i+3}\rangle_{3}|\phi^{-}\rangle_{4}\\\nonumber
|\Psi_{5,i}\rangle=|\psi^{+}\rangle_{1}|\psi^{-}\rangle_{2}|\omega^{i}\rangle_{3}|\psi^{+}\rangle_{4}
+|\psi^{-}\rangle_{1}|\psi^{+}\rangle_{2}|\omega^{i+1}\rangle_{3}|\psi^{-}\rangle_{4}
+|\phi^{+}\rangle_{1}|\phi^{-}\rangle_{2}|\omega^{i+2}\rangle_{3}|\phi^{+}\rangle_{4}
+|\phi^{-}\rangle_{1}|\phi^{+}\rangle_{2}|\omega^{i+3}\rangle_{3}|\phi^{-}\rangle_{4}\\\nonumber
|\Psi_{6,i}\rangle=|\psi^{+}\rangle_{1}|\psi^{-}\rangle_{2}|\omega^{i}\rangle_{3}|\psi^{+}\rangle_{4}
+|\psi^{-}\rangle_{1}|\psi^{+}\rangle_{2}|\omega^{i+1}\rangle_{3}|\psi^{-}\rangle_{4}
-|\phi^{+}\rangle_{1}|\phi^{-}\rangle_{2}|\omega^{i+2}\rangle_{3}|\phi^{+}\rangle_{4}
-|\phi^{-}\rangle_{1}|\phi^{+}\rangle_{2}|\omega^{i+3}\rangle_{3}|\phi^{-}\rangle_{4}\\\nonumber
|\Psi_{7,i}\rangle=|\psi^{+}\rangle_{1}|\psi^{-}\rangle_{2}|\omega^{i}\rangle_{3}|\psi^{+}\rangle_{4}
-|\psi^{-}\rangle_{1}|\psi^{+}\rangle_{2}|\omega^{i+1}\rangle_{3}|\psi^{-}\rangle_{4}
-|\phi^{+}\rangle_{1}|\phi^{-}\rangle_{2}|\omega^{i+2}\rangle_{3}|\phi^{+}\rangle_{4}
+|\phi^{-}\rangle_{1}|\phi^{+}\rangle_{2}|\omega^{i+3}\rangle_{3}|\phi^{-}\rangle_{4}\\\nonumber
|\Psi_{8,i}\rangle=|\psi^{+}\rangle_{1}|\psi^{-}\rangle_{2}|\omega^{i}\rangle_{3}|\psi^{+}\rangle_{4}
-|\psi^{-}\rangle_{1}|\psi^{+}\rangle_{2}|\omega^{i+1}\rangle_{3}|\psi^{-}\rangle_{4}
+|\phi^{+}\rangle_{1}|\phi^{-}\rangle_{2}|\omega^{i+2}\rangle_{3}|\phi^{+}\rangle_{4}
-|\phi^{-}\rangle_{1}|\phi^{+}\rangle_{2}|\omega^{i+3}\rangle_{3}|\phi^{-}\rangle_{4}\\\nonumber
|\Psi_{9,i}\rangle=|\psi^{+}\rangle_{1}|\phi^{+}\rangle_{2}|\omega^{i}\rangle_{3}|\psi^{+}\rangle_{4}
+|\psi^{-}\rangle_{1}|\phi^{-}\rangle_{2}|\omega^{i+1}\rangle_{3}|\psi^{-}\rangle_{4}
+|\phi^{+}\rangle_{1}|\psi^{+}\rangle_{2}|\omega^{i+2}\rangle_{3}|\phi^{+}\rangle_{4}
+|\phi^{-}\rangle_{1}|\psi^{-}\rangle_{2}|\omega^{i+3}\rangle_{3}|\phi^{-}\rangle_{4}\\\nonumber
|\Psi_{10,i}\rangle=|\psi^{+}\rangle_{1}|\phi^{+}\rangle_{2}|\omega^{i}\rangle_{3}|\psi^{+}\rangle_{4}
+|\psi^{-}\rangle_{1}|\phi^{-}\rangle_{2}|\omega^{i+1}\rangle_{3}|\psi^{-}\rangle_{4}
-|\phi^{+}\rangle_{1}|\psi^{+}\rangle_{2}|\omega^{i+2}\rangle_{3}|\phi^{+}\rangle_{4}
-|\phi^{-}\rangle_{1}|\psi^{-}\rangle_{2}|\omega^{i+3}\rangle_{3}|\phi^{-}\rangle_{4}\\\nonumber
|\Psi_{11,i}\rangle=|\psi^{+}\rangle_{1}|\phi^{+}\rangle_{2}|\omega^{i}\rangle_{3}|\psi^{+}\rangle_{4}
-|\psi^{-}\rangle_{1}|\phi^{-}\rangle_{2}|\omega^{i+1}\rangle_{3}|\psi^{-}\rangle_{4}
-|\phi^{+}\rangle_{1}|\psi^{+}\rangle_{2}|\omega^{i+2}\rangle_{3}|\phi^{+}\rangle_{4}
+|\phi^{-}\rangle_{1}|\psi^{-}\rangle_{2}|\omega^{i+3}\rangle_{3}|\phi^{-}\rangle_{4}\\\nonumber
|\Psi_{12,i}\rangle=|\psi^{+}\rangle_{1}|\phi^{+}\rangle_{2}|\omega^{i}\rangle_{3}|\psi^{+}\rangle_{4}
-|\psi^{-}\rangle_{1}|\phi^{-}\rangle_{2}|\omega^{i+1}\rangle_{3}|\psi^{-}\rangle_{4}
+|\phi^{+}\rangle_{1}|\psi^{+}\rangle_{2}|\omega^{i+2}\rangle_{3}|\phi^{+}\rangle_{4}
-|\phi^{-}\rangle_{1}|\psi^{-}\rangle_{2}|\omega^{i+3}\rangle_{3}|\phi^{-}\rangle_{4}\\\nonumber
|\Psi_{13,i}\rangle=|\psi^{+}\rangle_{1}|\phi^{-}\rangle_{2}|\omega^{i}\rangle_{3}|\psi^{+}\rangle_{4}
+|\psi^{-}\rangle_{1}|\phi^{+}\rangle_{2}|\omega^{i+1}\rangle_{3}|\psi^{-}\rangle_{4}
+|\phi^{+}\rangle_{1}|\psi^{-}\rangle_{2}|\omega^{i+2}\rangle_{3}|\phi^{+}\rangle_{4}
+|\phi^{-}\rangle_{1}|\psi^{+}\rangle_{2}|\omega^{i+3}\rangle_{3}|\phi^{-}\rangle_{4}\\\nonumber
|\Psi_{14,i}\rangle=|\psi^{+}\rangle_{1}|\phi^{-}\rangle_{2}|\omega^{i}\rangle_{3}|\psi^{+}\rangle_{4}
+|\psi^{-}\rangle_{1}|\phi^{+}\rangle_{2}|\omega^{i+1}\rangle_{3}|\psi^{-}\rangle_{4}
-|\phi^{+}\rangle_{1}|\psi^{-}\rangle_{2}|\omega^{i+2}\rangle_{3}|\phi^{+}\rangle_{4}
-|\phi^{-}\rangle_{1}|\psi^{+}\rangle_{2}|\omega^{i+3}\rangle_{3}|\phi^{-}\rangle_{4}\\\nonumber
|\Psi_{15,i}\rangle=|\psi^{+}\rangle_{1}|\phi^{-}\rangle_{2}|\omega^{i}\rangle_{3}|\psi^{+}\rangle_{4}
-|\psi^{-}\rangle_{1}|\phi^{+}\rangle_{2}|\omega^{i+1}\rangle_{3}|\psi^{-}\rangle_{4}
-|\phi^{+}\rangle_{1}|\psi^{-}\rangle_{2}|\omega^{i+2}\rangle_{3}|\phi^{+}\rangle_{4}
+|\phi^{-}\rangle_{1}|\psi^{+}\rangle_{2}|\omega^{i+3}\rangle_{3}|\phi^{-}\rangle_{4}\\\nonumber
|\Psi_{16,i}\rangle=|\psi^{+}\rangle_{1}|\phi^{-}\rangle_{2}|\omega^{i}\rangle_{3}|\psi^{+}\rangle_{4}
-|\psi^{-}\rangle_{1}|\phi^{+}\rangle_{2}|\omega^{i+1}\rangle_{3}|\psi^{-}\rangle_{4}
+|\phi^{+}\rangle_{1}|\psi^{-}\rangle_{2}|\omega^{i+2}\rangle_{3}|\phi^{+}\rangle_{4}
-|\phi^{-}\rangle_{1}|\psi^{+}\rangle_{2}|\omega^{i+3}\rangle_{3}|\phi^{-}\rangle_{4}\\\nonumber
\end{eqnarray}
\end{widetext}
In the above states, $i=0,...,3$ and $|\omega^{i}\rangle$ are
define as:
\begin{eqnarray*}
|\omega^{0,1}\rangle=|\Psi^{\pm}\rangle
\hspace{.5cm}|\omega^{2,3}\rangle=|\Phi^{\pm}\rangle
\end{eqnarray*}
the subscript $i$ is calculated in the $mod \hspace{.2cm} 4$. In
the above states, the first two Bell states of $1,2$ belongs to
Alice and the second two Bell states of $3,4$ belongs to Bob which
are represented by subscriptions in the Bell states respectively.

Similar to the previous case,  Alice and Bob share only one of the
above states (for example, $|\Psi_{1,0}\rangle$) and Alice can
transform this state to one of the above states (one quarter of
the total Hilbert space dimension) by local operation on the her
qubits.

They had previously compromised that if Alice states received by
Bob were in the time intervals $\delta t_{0}=t_{1}-t_{0}$ or
$\delta t_{1}=t_{2}-t_{1}$ or $\delta t_{2}=t_{2}-t_{3}$ or
$\delta t_{3}=t_{4}-t_{3}$, Bob would operate $I^{B}\otimes I^{B}$
or $\sigma_{z}^{B}\otimes I^{B}$ or $I^{B}\otimes \sigma_{x}^{B}$
or $\sigma_{z}^{B}\otimes \sigma_{x}^{B}$ on his qubits
respectively \cite{Ano2}. By using this protocol, Alice and Bob
get access to all of the Hilbert space states.

Similar to the previous cases, Karus operators $A_{i,j,k,l}$ ,
satisfy $\sum_{i,j,k,l}A_{i,j,k,l}A_{i,j,k,l}^{\dagger}=1$, and in
the presence of partial memory, the action of Pauli channels on
the $\rho$ is described by:
\begin{widetext}
\begin{eqnarray*}
A_{i,j,k,l}(\mu)=\sqrt{p_{i}[(1-\mu)p_{j}p_{k}p_{l}+\mu\delta_{i,j}\delta_{j,k}\delta_{k,l}]}
\sigma_{i}\otimes\sigma_{j}\otimes\sigma_{k}\otimes\sigma_{l}\otimes
I^{\otimes4}
\end{eqnarray*}
\end{widetext}
For simplicity, we consider only two types of sending particles:
i) The two Bell states in the Alice's hand (subscribed by $1,2$ in
the states (\ref{states})) are sent at the same time
($\tau_{lap}\ll\tau_{fluc}$ for each pair). ii) The two Bell
states are sent with a time delay ($\tau_{fluc}\ll \tau_{lap}$ for
each pair). Although, we can consider the general case of the
transmission of quantum states, but its explicit form is very
complicated and doesn't clarify any physical properties. With a
similar approach to the one we described in the previous case, the
mutual information for the entanglement-assisted systems can be
calculated. If Alice sends the density matrix $\rho$ through a
quantum channel, the corresponding output state is given by the
map:
\begin{eqnarray*}
\rho \longrightarrow
\varepsilon(\rho)=\sum_{i,j,k,l=0}^{3}A_{i,j,k,l}(\mu)\rho
A_{i,j,k,l}^{\dagger}(\mu)
\end{eqnarray*}
For channels with partial memory output, the density matrix can be
written as following:
\begin{widetext}
\begin{eqnarray*}
\varepsilon(\rho)&=&(1-\mu)\sum_{i,j,k,l=0}^{3}p_{i}p_{j}p_{k}p_{l}
(\sigma_{i}\otimes\sigma_{j}\otimes\sigma_{k}\otimes\sigma_{l}\otimes
I^{\otimes4})\rho
(\sigma_{i}\otimes\sigma_{j}\otimes\sigma_{k}\otimes\sigma_{l}\otimes
I^{\otimes4})\\\nonumber
&&+\mu\sum_{i=0}^{3}p_{i}(\sigma_{i}\otimes\sigma_{i}\otimes\sigma_{i}\otimes\sigma_{i}\otimes
I^{\otimes4})\rho
(\sigma_{i}\otimes\sigma_{i}\otimes\sigma_{i}\otimes\sigma_{i}\otimes
I^{\otimes4})
\end{eqnarray*}
\end{widetext}
After a similar calculation to the previous case, we derive the
mutual information of the quantum channel, with correlated noise,
given by:
\begin{eqnarray}
I_{2}^{AB, e-a}(\varepsilon, \pi_{i})=
8-S(\varepsilon^{A}\otimes\emph{I}^{B}(\rho^{AB}))
\end{eqnarray}
The von Neumann entropy of the $\rho^{AB}$ state, transformed
under the action of Pauli channels, is given by:

\begin{eqnarray}
S(\varepsilon^{A}\otimes\emph{I}^{B}(\rho^{AB}))=-\sum_{i}\lambda_{i}\log_{2}\lambda_{i}
\end{eqnarray}
where $\lambda_{i}$ are the eigenvalues of the transformed
$\rho^{AB}$ state with the explicit forms:
\begin{eqnarray*}
\lambda_{1}&=&(1-\mu)(p_{0}^{4}+p_{z}^{4}+p_{x}^{4}+p_{y}^{4})+\mu\\\nonumber
\lambda_{2,3,4}&=&(1-\mu)[2(p_{0}p_{z})^{2}+2(p_{x}p_{y})^{2}]\\\nonumber
\lambda_{4,6,7}&=&(1-\mu)[2(p_{0}p_{x})^{2}+2(p_{y}p_{z})^{2}]\\\nonumber
\lambda_{8,9,10}&=&(1-\mu)[2(p_{0}p_{y})^{2}+2(p_{x}p_{z})^{2}]\\\nonumber
\lambda_{11,...,14}&=&(1-\mu)[p_{0}p_{z}(p_{0}^{2}+p_{z}^{2})+p_{x}p_{y}(p_{x}^{2}+p_{y}^{2})]\\\nonumber
\lambda_{15,...,26}&=&(1-\mu)[p_{0}p_{z}(p_{x}^{2}+p_{y}^{2})+p_{x}p_{y}(p_{0}^{2}+p_{z}^{2})]\\\nonumber
\lambda_{27,...,30}&=&(1-\mu)[p_{0}p_{x}(p_{0}^{2}+p_{x}^{2})+p_{y}p_{z}(p_{y}^{2}+p_{z}^{2})]\\\nonumber
\lambda_{31,...,38}&=&(1-\mu)[p_{0}p_{x}(p_{y}^{2}+p_{z}^{2})+p_{y}p_{z}(p_{0}^{2}+p_{x}^{2})]\\\nonumber
\lambda_{39,...,42}&=&(1-\mu)[p_{0}p_{y}(p_{x}^{2}+p_{z}^{2})+p_{x}p_{z}(p_{0}^{2}+p_{y}^{2})]\\\nonumber
\lambda_{43,...,46}&=&(1-\mu)[p_{0}p_{y}(p_{0}^{2}+p_{y}^{2})+p_{x}p_{z}(p_{x}^{2}+p_{z}^{2})]\\\nonumber
\lambda_{47,...,52}&=&(1-\mu)[4p_{0}p_{x}p_{y}p_{z}]\\\nonumber
\lambda_{53,...,60}&=&(1-\mu)[(p_{0}p_{z}+p_{x}p_{y})(p_{0}p_{x}+p_{y}p_{z})]\\\nonumber
\lambda_{61,...,64}&=&(1-\mu)[(p_{0}p_{z}+p_{x}p_{y})(p_{0}p_{y}+p_{x}p_{z})]\\\nonumber
\end{eqnarray*}
Here we have four sets of equal eigenvalues (corresponding to each
four sets of states, in which every set has $64$ elements). These
states don't mix with each other through the interaction with
environment.

\begin{figure}
\centering
\includegraphics[height=6.5cm,width=8cm]{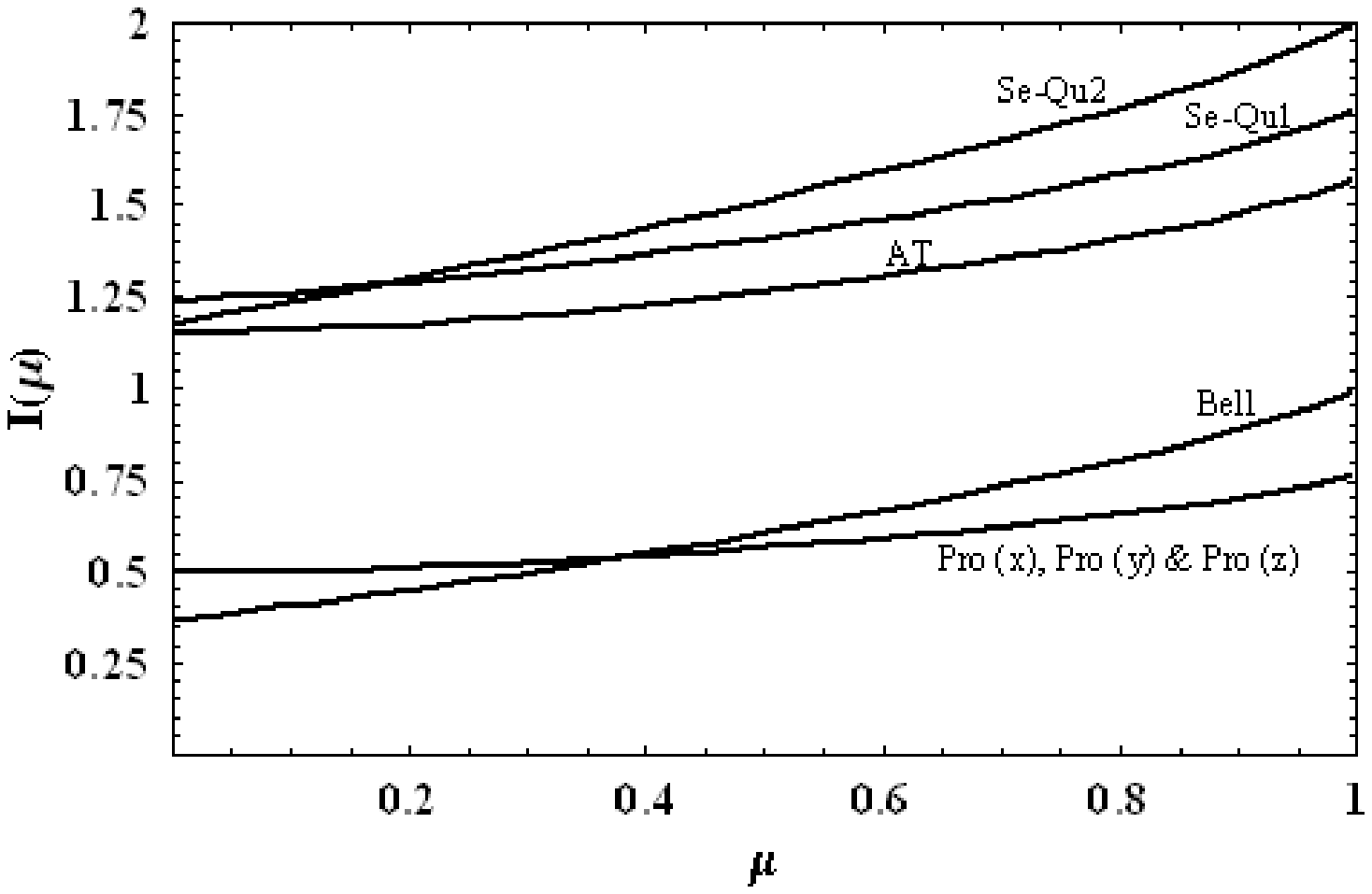}
\caption{  Mutual information $I(\mu)$ for product state in the
$x$ basis (Pro(x)), $y$ basis (Pro(y)) and $z$ basis (Pro(z)),
maximally entanglement state (Bell), entanglement-assisted (AT),
first semi-quantum approach (Se-Qu1) and second semi-quantum
approach (Se-Qu2) versus the memory coefficient $\mu$ for
depolarizing channel, with $p_{0}=0.85$ and $p_{i}=0.05,i=x,y,z$.
The capacities are normalized with respect to the number of
channel uses. Value of memory threshold of (Se-Qu1) and (Se-Qu2)
for our choice of the probability of errors channel is
$\mu_{t}=0.171$.}\label{Dep}
\end{figure}

\begin{figure}
\centering
\includegraphics[height=7cm,width=8cm]{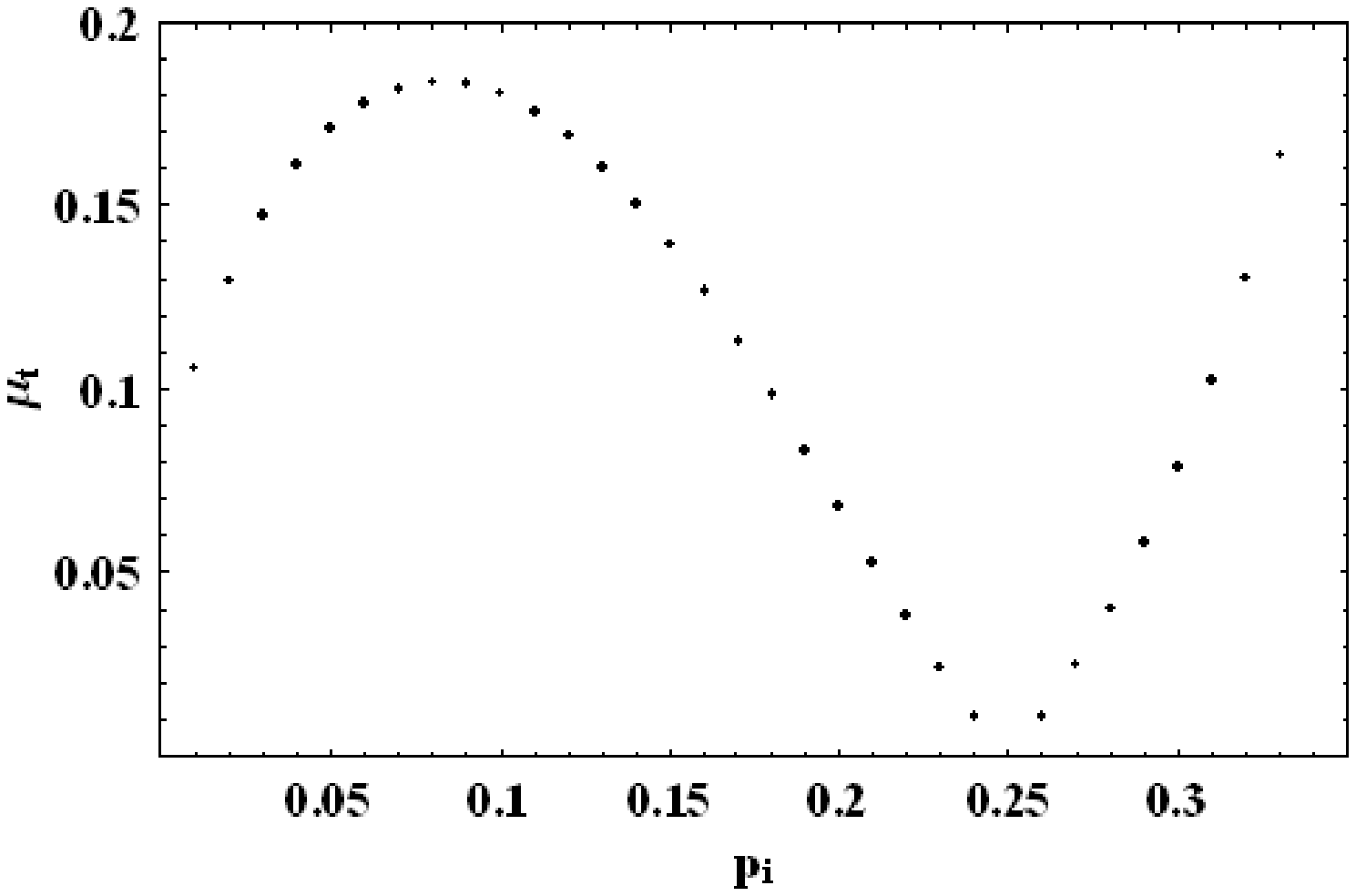}
\caption{  Memory threshold $\mu_{t}$ as a function of the
probability of errors $p_{i},\hspace{.2cm}i=x,y,z$ in the
depolarizing channel for first semi-quantum approach (Se-Qu1) and
second semi-quantum approach (Se-Qu2).} \label{Mem}
\end{figure}
In the following, we would like to consider some examples of Pauli
channels, both symmetric and asymmetric, and to work out their
entanglement-assisted classical capacity. For depolarizing
channels which are symmetric types of Pauli channel, for which
$p_{0}=1-p$ and $p_{i}=p/3,i=x,y,z$. In Fig.(\ref{Dep}), we plot
the mutual information of quantum channel
$I^{AB}(\varepsilon(\rho))$ and the mutual information
corresponding to entanglement-assisted dense coding system $I^{AB,
e-a}(\varepsilon(\rho))$ of the depolarization channel versus its
memory coefficient $\mu$. Fig.(\ref{Dep}) shows that the use of
semi-quantum approach enhances mutual information of the quantum
channel, if one prior entanglement is shared by Alice and Bob.
That has better capacity over both products, entangled states and
entanglement-assisted coding for all values of $\mu$. Similar to
earlier works \cite{Mac}, there exists another memory threshold
for two semi-quantum approaches. Explicit form of the memory
threshold is a very complicated relation, but numerical
calculation is shown in the Fig.(\ref{Mem}) for $0<p_{i}<1/3$.

In Fig.(\ref{Mem}) we see that as expected in the high error
channels, $p_{0}=p_{i}=p/3,i=x,y,z$, the memory threshold is equal
to zero, i.e. $\mu_{t}=0$. In other words, in the channels with
high errors, any output density matrix can be transformed to the
following form:
\begin{eqnarray}
\varepsilon (\rho)= (1-\mu)\frac{1}{a^{b}}I^{\otimes b}+\mu \sigma
\end{eqnarray}
With $a=b=4$ for states (\ref{states}) and $a=4,b=2$ for states
(\ref{Sem-Qun1}). Hence, $Tr\sigma=1, Tr\varepsilon (\rho)=1$.
Optimal mutual information is obtained by minimizing the output
entropy, and, for this, we must have a pure state at the output
channel. The optimization of the the mutual information can be
achieved by going to an appropriate bases that diagonaliyez
$\sigma$. If we assume that $\sigma$ has $k$ none-zero diagonal
elements, then, the entropy is given by:
\begin{eqnarray*}
S(\varepsilon (\rho))&=&k\{(1-\mu)a^{-b}+\mu\frac{1}{k}
\}\log_{2}\{(1-\mu)a^{-b}+\mu\frac{1}{k}\}\nonumber\\
&&+(a^{b}-k)\{(1-\mu)a^{-b}\}\log_{2}\{1-\mu)a^{-b}\}
\end{eqnarray*}
Minimum value of the above relation can be obtained for $k=1$. In
the other words, $\sigma$ must be a pure state, and this happens
for the input states of (\ref{states}).

On other hand, the maximum of memory threshold is give by $\mu_{t}
^{Max}=0.185$. Thus, for channels with $\mu\geq\mu_{t} ^{Max}$,
the use of (Se-Qu2) approach is more efficient.

\begin{figure}
\centering
\includegraphics[height=6.5cm,width=8cm]{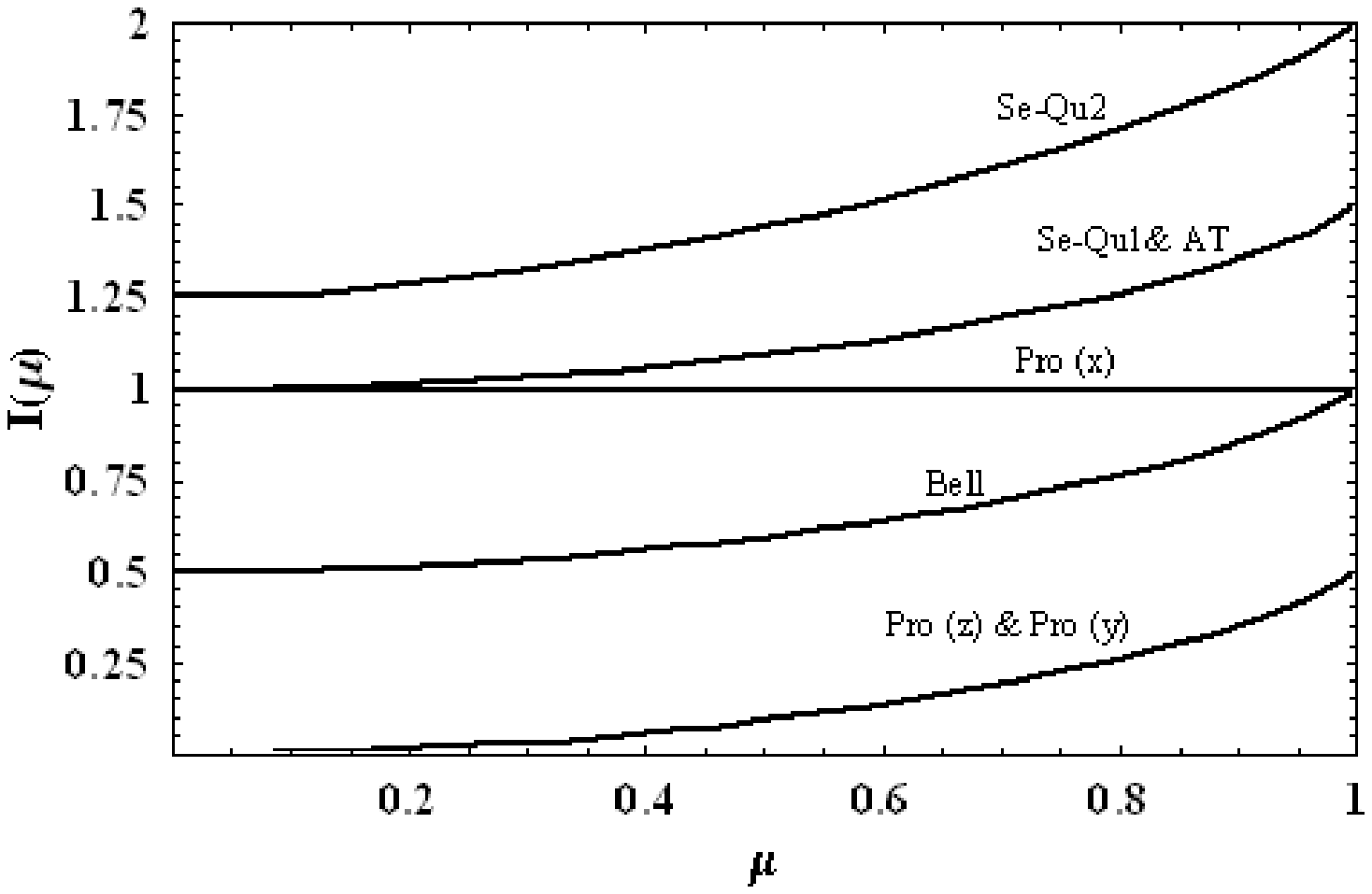}
\caption{  Mutual information $I(\mu)$ for the product state in
the $x$ basis (Pro(x)), $y$ basis (Pro(y)) and $z$ basis (Pro(z)),
maximally entanglement state (Bell), entanglement-assisted (AT),
first semi-quantum approach (Se-Qu1) and second semi-quantum
approach (Se-Qu2) versus the memory coefficient $\mu$ for flip
channel (bit flip), with $p_{0}=0.5$ and $p_{x}=0.5$. The
capacities are normalized with respect to the number of channel
uses.} \label{Bit}
\end{figure}
In the following, we consider some examples of asymmetric Pauli
channels \cite{nil}. The noise introduced by them is of three
types: namely, bit flip, phase flip and bit-phase flip.
Probability distribution for flip channels is given by $i,j=0,f$,
with probabilities  $p_{0}=1-p$ and $p_{f}=p$. Here $f=x,y$ and
$z$ for bit flip, phase flip and bit-phase flip channels,
respectively. In Fig.(\ref{Bit}), we plot the mutual information
of $I^{AB}(\varepsilon(\rho)$ and the mutual information
corresponding to entanglement-assisted dense coding system $I^{AB,
e-a}(\varepsilon(\rho)$ of the bit flip channel versus its memory
coefficient $\mu$. Fig.(\ref{Bit}) shows that the use of four
particles, semi-quantum dense coding, approach enhances the
capacity of quantum channel and the usual entanglement-assisted
coding and two particle semi-quantum dense coding have the same
plot for all values of $\mu$. This shows that in the asymmetric
Pauli channels, optimality in the unshared entangled states
strictly depends on the noise of the channel. For example, in the
bit flip channels, it is more appropriate to use $x$ basis.

\begin{figure}
\centering
\includegraphics[height=6.5cm,width=8cm]{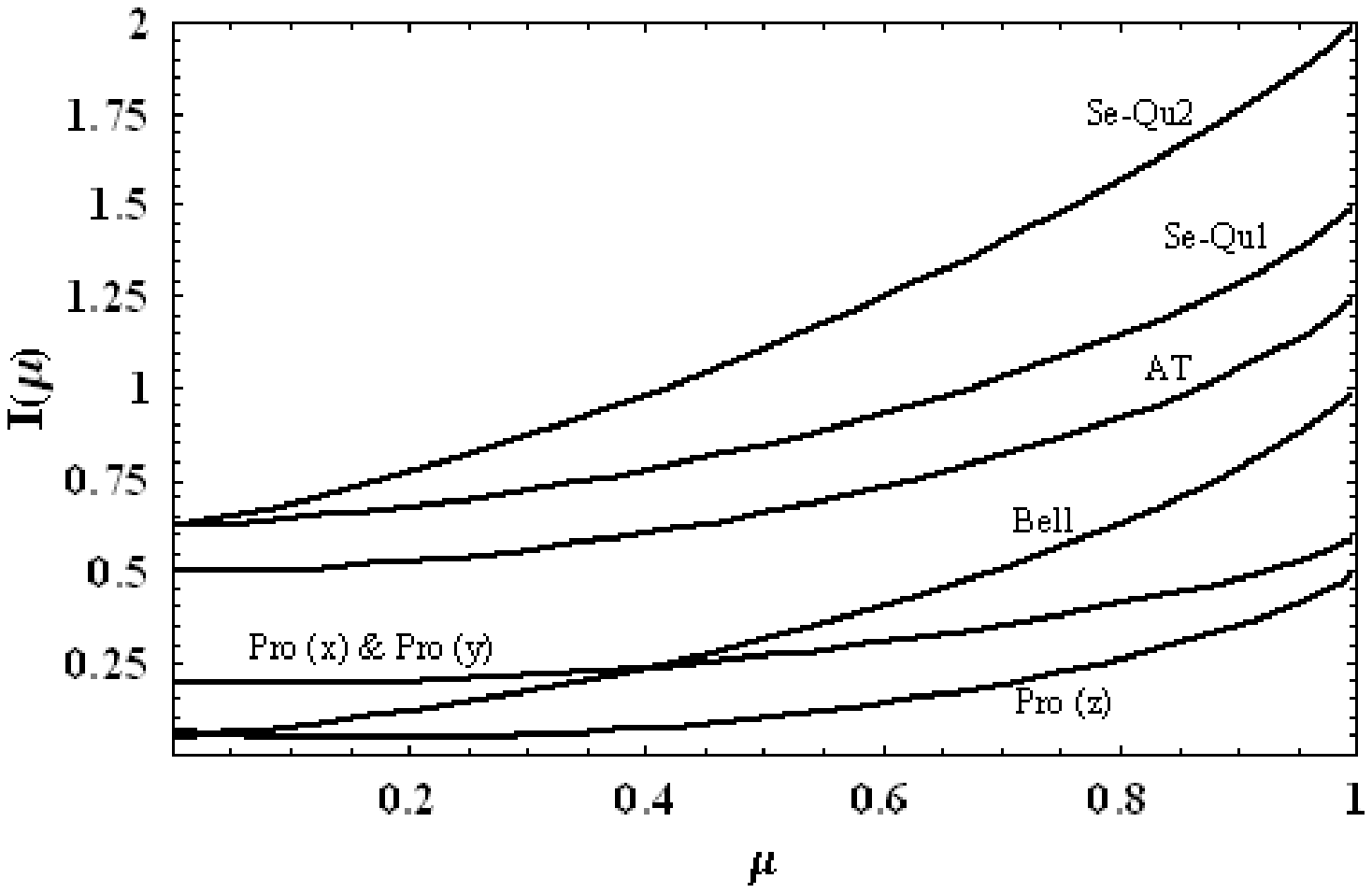}
\caption{  Mutual information $I(\mu)$ for product state in the
$x$ basis (Pro(x)), $y$ basis (Pro(y)) and $z$ basis (Pro(z)),
maximally entanglement state (Bell), entanglement-assisted (AT),
first semi-quantum approach (Se-Qu1) and second semi-quantum
approach (Se-Qu2) versus the memory coefficient $\mu$ for
two-Pauli channel, with $p_{0}=0.5$ and
$p_{i}=0.25,\hspace{.2cm}i=x,y$. The capacities are normalized
with respect to the number of channel uses. Value of memory
threshold of (Pro(x),Pro(y)) state and (Bell) for our choice of
probability of errors channel is $\mu_{t}=0.409$.} \label{Two}
\end{figure}

\begin{figure}
\centering
\includegraphics[height=6.52cm,width=8cm]{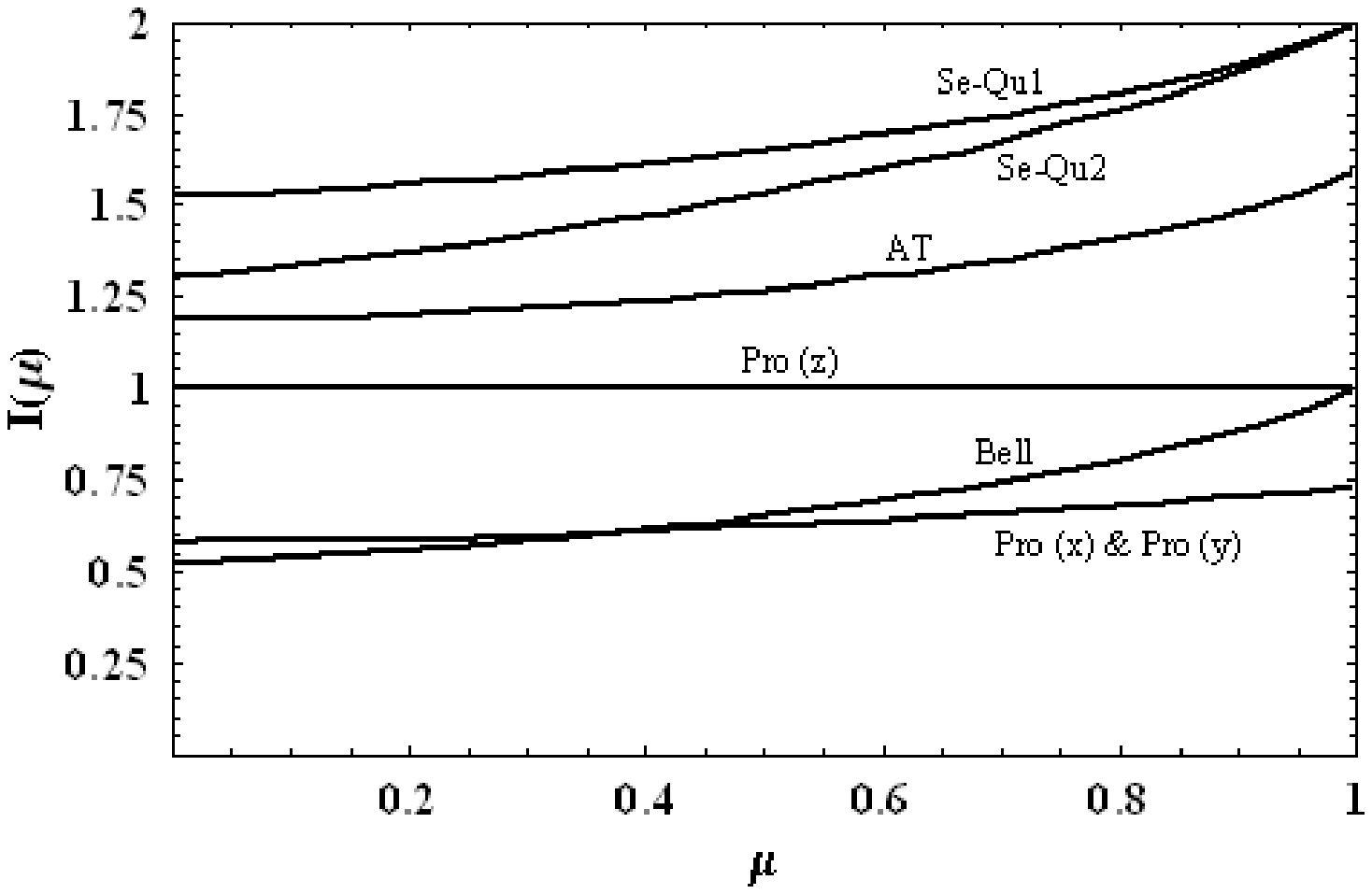}
\caption{  Mutual information $I(\mu)$ for product state in the
$x$ basis (Pro(x)), $y$ basis (Pro(y)) and $z$ basis (Pro(z)),
maximally entanglement state (Bell), entanglement-assisted (AT),
first semi-quantum approach (Se-Qu1) and second semi-quantum
approach (Se-Qu2) versus the memory coefficient $\mu$ for phase
damping channel, with $p_{0}=0.75$ and $p_{z}=0.25$. The
capacities are normalized with respect to the number of channel
uses.} \label{Phase}
\end{figure}
At another stage, we consider the two-Pauli channels \cite{nil}.
The probability distribution for the two-Pauli channels is given
by $i,j=0,x,y$, with probabilities $p_{0}=(1-p)$ and
$p_{x}=p_{y}=p/2$. In Fig.(\ref{Two}), we plot the mutual
information of $I^{AB}(\varepsilon(\rho)$ and the mutual
information corresponding to entanglement-assisted dense coding
system $I^{AB, e-a}(\varepsilon(\rho)$ of the two-Pauli channels,
versus its memory coefficient $\mu$. This shows that the use of
four particles semi-quantum dense coding approach enhances the
capacity of quantum channel for all values of $\mu$. Furthermore,
at the usual stage (unshared state) \cite{Mac}, use of product
states in the $x$ or $y$
 basis would be more efficient than in the $z$ basis and there exists a memory threshold where a higher amount of
 classical information is transmitted with entangled states.
 Explicit form of memory threshold is very complicated, the numerical value of memory threshold
 for the
 above error model is $\mu_{t}=0.409$.

Finally, we consider another type of asymmetric Pauli channels,
the so-called phase damping channel \cite{nil}. The probability
distribution for phase damping channels are given by $i,j=0,z$
with probabilities $p_{0}=(1-p/2)$ and $p_{z}=p/2$.
Fig.(\ref{Phase}) shows that for phase damping channels, the use
of two particles semi-quantum dense coding is more appropriate,
compared with the four particle approach to information
transmission, for all of values of $\mu$. This shows that in the
usual stage (previously unshared state) \cite{Mac}, the use of
product states in the $z$
 basis would be more efficient than in the $x$ or $y$ basis.

On the other hand, we show that $I_n({\cal E})$ is superadditive
in the presence of entanglement-assisted, i.e. we have
$I^{E}_{n+m}> I^{E}_{n} + I^{E}_{m}$ and therefore $C^{E}_n >
C^{E}_1$. At this stage, the classical capacity $C^{E}$ of the
channel is defined by:
\begin{eqnarray*}
C^{E}=\lim_{n\to \infty}  C^{E}_n\;. \label{capac}
\end{eqnarray*}
It has been shown (similar to what we plotted in the figures) that
the amount of reliable information which can be transmitted per
use of the channel, is given by \cite{Sch}:
\begin{eqnarray*}
C^{E}_n =\frac{1}{n}{\mbox{sup}}_{\cal E} I^{E}_n(\cal E)\ \ ,
\end{eqnarray*}



One of the main applications of the above extension of memorial
channels is the extension of the standard quantum cryptography
BB84 \cite{Benn} to protocols where the key is carried by quantum
states in a space of higher dimension, using two (or $d+1$)
mutually unbiased bases, which for high memorial channels have
very low error rates. This procedure ensures that any attempt by
any eavesdropper Eve to gain information on sender's state induces
errors in the transmission, which can be detected by the
legitimate parties \cite{Cerf}. On the other hand, if we are
interested in other quantum key distribution protocols (such as
the EPR protocol \cite{Eke}), we must encode a qubit in a
decoherence-free subspace of the collective noise for key
distribution \cite{Boi1}. If we are interested in the total
dimension of Hilbert space, we must revise  the EPR protocol for
this new approach. Another application of the above extension can
be quantum coding and quantum superdence coding at higher
dimensions. The errors in the memorial channels can be considered
as a subset of collective noise which are considered in the
decoherence-free subspace approach. Some experimental results
\cite{Ball} show that in some special cases the use of these
states are appropriate, because in the memorial channels we make
use of all maximally entangled states.

In conclusion, we have calculated a general expression for the
classical capacity of a quantum channel for product states,
maximally entangled states and entanglement-assisted coding in the
presence of memory. Hence, we have suggested another approach for
the transmission of information by using semi-quantum dense
coding. In this approach, we use an entangled pair that was
previously shared between the parties, and they compromised for a
transmission of quantum states through quantum channels in a
specific manner. Our results show that if noise in the consecutive
uses of the channels is assumed to be Markov-correlated quantum
noise, then, the use of semi-quantum approach to quantum dense
coding enhances classical capacity of quantum channels in various
types of Pauli channels.

We would like to thank M. Golshani for useful comments and
critical reading of the manuscript and S. Fallahi for his
helps.(This work was supported under the project: \emph{Entezar}).



\begin{references}
\bibitem {Ano1} There exists another approach to this protocol, where Alice and Bob
have two quantum channels that we call them first and second
channels. If Alice sends her qubits from the first (second)
channel, then, Bob operates on the his qubits by $I^{B}\otimes
I^{B}$ ($I^{B}\otimes \sigma_{x}^{B}$).
\bibitem {Ano2} There exists another approach to this protocol, where Alice and Bob
have four quantum channels that we call first, second, third and
fourth channel. If Alice sends her qubits from the first, second,
third and fourth channels, then, Bob operates on the fourth Bell
state by $I^{B}\otimes I^{B}$ or $\sigma_{z}^{B}\otimes I^{B}$ or
$I^{B}\otimes \sigma_{x}^{B}$ or $\sigma_{z}^{B}\otimes
\sigma_{x}^{B}$ respectively.
\end{references}
\end{document}